\documentclass[a4paper,preprint,times,5p,sort&compress]{elsarticle}
\usepackage{amsmath, amssymb}
\usepackage{multirow}
\usepackage{longtable}
\usepackage{color}
\usepackage{hyperref}
\usepackage[normalem]{ulem}  % \sout{old text} for strikeout
\begin{document}
 \begin{frontmatter}

  \title{Criticality of the net-baryon number probability distribution at
  finite density}
  \date{\today}
  \author[fias,yitp,wroclaw]{Kenji Morita\corref{cor}}
  \ead{morita@fias.uni-frankfurt.de}
  \cortext[cor]{Corresponding author}
  \address[fias]{Frankfurt Institute for Advanced
  Studies,Ruth-Moufang-Strasse 1, D-60438, Frankfurt am Main, Germany}
  \address[yitp]{Yukawa Institute for Theoretical Physics, Kyoto University,
  Kyoto 606-8502, Japan}
  \author[gsi]{Bengt Friman}
  \address[gsi]{GSI, Helmholzzentrum f\"{u}r Schwerionenforschung,
  Planckstrasse 1, D-64291 Darmstadt, Germany}
  \author[wroclaw,emmi]{Krzysztof Redlich}
  \address[wroclaw]{Institute of Theoretical Physics, University of Wroclaw,
  PL-50204 Wroc\l aw, Poland}
  \address[emmi]{Extreme Matter Institute EMMI, GSI,
  Planckstrasse 1, D-64291 Darmstadt, Germany}

 \begin{abstract}
  We compute the probability distribution $P(N)$ of the net-baryon
  number at finite temperature and
  quark-chemical potential, $\mu$, at a physical value of the pion mass  in the quark-meson model within the
  functional renormalization group scheme. For $\mu/T<1$,
  the model exhibits the chiral crossover transition which belongs to the universality class of the $O(4)$
  spin system in three dimensions.
  We explore the influence of the chiral crossover transition on the properties of the
  net baryon number probability distribution, $P(N)$. By
  considering ratios of $P(N)$ to the Skellam function, with
  the same mean and variance, we unravel the characteristic features
  of the distribution that are related to $O(4)$ criticality at
  the chiral crossover transition. We explore the corresponding ratios for data obtained
  at RHIC by the STAR Collaboration and discuss their
  implications. We also examine $O(4)$ criticality in the context
  of binomial and negative-binomial distributions for the net proton number.
 \end{abstract}

  \begin{keyword}
   Chiral phase transition, Conserved charge fluctuations, Probability
   distribution, Heavy ion collisions
  \end{keyword}
\end{frontmatter}

%\linenumbers
%\Large

\section{Introduction}
Fluctuations of conserved charges are promising observables for exploring
critical phenomena in relativistic heavy ion collisions
\cite{stephanov99:_event_by_event_fluct_in,asakawa00:_fluct_probes_of_quark_decon,jeon00:_charg_partic_ratio_fluct_as,hatta03:_proton_number_fluct_as_signal}.
A particular role is attributed to higher order cumulants of the
net baryon number and electric charge fluctuations,  which in a QCD
medium, can be negative near the chiral transition
\cite{skokov10:_meson_fluct_and_therm_of,stephanov09:_non_gauss_fluct_near_qcd_critic_point,stephanov11:_sign_of_kurtos_near_qcd_critic_point}.

At physical values of quark masses, the  phase transition in QCD is expected to change from a
crossover transition at small values of the baryon chemical potential to a first-order transition
at large net baryon densities.  The  first-order chiral phase transition, if it exists, then begins
in a second-order critical point, the critical end point (CEP)
\cite{asakawa89:_chiral_restor_at_finit_densit_and_temper}.
Owing to the divergent correlation length at the
CEP \cite{stephanov11:_sign_of_kurtos_near_qcd_critic_point}, and the
spinodal phase separation  in a non-equlibrium first-order transition
\cite{sasaki08:_chiral_phase_trans_in_presen}, one expects
large fluctuations of the net-baryon number in heavy ion collisions,
at beam energies where the system passes through the first-order phase boundary
or close to the CEP.

The conjectured existence of a CEP in the QCD phase diagram
has so far not been  confirmed  by  lattice QCD calculations
(LQCD) \cite{forcrand09:_simul_qcd_at_finit_densit,forcrand07:_n_qcd}.
At small values of the quark chemical potential, $\mu_q /T < 1$,
LQCD exhibits a chiral crossover transition. There are indications that,
in the chiral limit for light quarks, the QCD transition belongs to the universality class
of 3-dimensional $O(4)$ spin systems
\cite{pisarski84:_remar_chiral_trans_in_chrom,ejiri09:_magnet_equat_of_state_in_flavor_qcd}.
Thus, a promising approach for probing the phase boundary
in heavy ion collisions, is to explore the fluctuations of the chiral phase
transition, assuming $O(4)$ criticality. Owing to the proximity of the chemical freeze-out to the chiral
crossover at small values of the baryonic chemical potential, one may expect that
the critical fluctuations are reflected in the data on conserved charges
\cite{karsch11:_probin_freez_out_condit_in}. A baseline for the
cumulants of charge fluctuations is provided by the hadron
resonance gas (HRG) model, which reproduces the particle yields at
chemical freeze-out in heavy ion collisions \cite{Statmodelreview_QGP3},
as well as the LQCD equation of state in the hadronic phase
\cite{borsanyi10:_qcd,huovinen10:_qcd}.

At the CEP, which is expected to belong to 3d $Z(2)$ universality class,
the second and higher order cumulants diverge. By contrast, at a chiral phase transition
belonging to the $O(4)$ universality class, at vanishing baryon chemical potential,
low-order cumulants remain finite, while the sixth and higher order cumulants
diverge\footnote{For $\mu\neq 0$, diverging cumulants appear already
at third order.}. For non-zero quark masses, the divergences are
replaced by a rapid variation of the cumulants near the crossover
temperature, including changes of sign~\cite{friman11:_fluct_as_probe_of_qcd}.

The fluctuations of the net-baryon number, more precisely of the
net-proton number,  were measured in heavy ion collisions by STAR
collaboration at RHIC
\cite{aggarwal10:_higher_momen_of_net_proton,luo13:_searc_for_qcd_critic_point,braun-munzinger11:_net_proton_probab_distr_in,STAR_pn_2013}.
Data on the mean ($M$), variance ($\sigma$), skewness ($S$) and kurtosis ($\kappa$)
of the net-proton number were obtained in a broad energy range and for
different centralities. These observables are linked to the
cumulants   $\chi_n$  of the net-baryon number, and are
accordingly modified by the critical chiral dynamics.

The most recent STAR data  show that, while the ratio $\sigma^2/M=\chi_2/\chi_1$
is consistent with the HRG result in central collisions, differences are found in the products  $S\sigma=\chi_3/\chi_1$  and
$\kappa\sigma^2=\chi_4/\chi_2$. The deviations in the latter are small
at the  top  RHIC energy, increase with the order of the cumulants
at fixed collision energy, and show a non-monotonic dependence on the energy
with a maximum at  $\sqrt{s_{NN}} \simeq 19$ GeV.
In the $O(4)$ universality class
one expects\footnote{Although the results of \cite{morita12:_baryon_number_probab_distr_near} were obtained in the chiral limit,
it is plausible that they remain valid also for physical values of the quark masses.} $\chi_4/\chi_2<1$, while in $Z(2)$ this ratio is expected to be larger than
unity \cite{morita12:_baryon_number_probab_distr_near}. Thus, the systematics
of the ratios of cumulants in central Au-Au
collisions, as measured by STAR \cite{STAR_pn_2013},  indicate that the observed deviations
of  the net proton number fluctuations from the HRG values may be attributed to $O(4)$
criticality at  the  phase boundary, at least for
$\sqrt{s_{NN}}\geq 19$ GeV.

The cumulants of a conserved charge are given by appropriate
combinations of moments of the corresponding
probability distribution. Thus, the behavior of cumulants near criticality
must be reflected in the properties of the probability distribution.
Moreover, it is expected that the critical behavior of the probability distribution
depend on the universality class.
Indeed, we have recently shown, that the structure of the probability
distribution for the net baryon number depends on the  properties  of the
critical chiral fluctuations
\cite{morita12:_baryon_number_probab_distr_near,morita13:_net}.
In particular, we have argued,  that  at vanishing chemical potential,
the residual $O(4)$ critical fluctuation at physical pion mass leads to
narrowing of the probability distribution relative to the Skellam
function. This corresponds to a negative structure of the
sixth order cumulant at the chiral crossover transition~\cite{morita13:_net}.

In this paper,  we extend our previous studies to non-zero chemical
potential and propose  a method  for  identifying the characteristic properties of the
net baryon probability distribution, which are responsible for
the critical behavior of the cumulants at the chiral transition.
We apply this method to the net proton probability distributions
obtained  by the STAR collaboration in central Au-Au collisions at
$\sqrt{s_{NN}} \geq 19$ GeV.  We also critically examine the question
whether $O(4)$ criticality can be captured by assuming that the baryon and antibaryon
mutliplicities are described by binomial or negative binomial distributions.

In this paper, we show that the (suitably rescaled) ratio of the net baryon
probability distribution to the  corresponding Skellam function, reveals the critical narrowing of the probability
distribution, which is characteristic for the $O(4)$ scaling.

\section{The net-baryon number probability distribution}
In the grand canonical ensemble specified by temperature $T$, subvolume $V$
and chemical potential $\mu$, the probability distribution for the
conserved charge $N$,  is given by
\begin{equation}
 P(N;T,V,\mu) = \frac{Z(T,V,N)e^{\mu N/T}}{\mathcal{Z}(T,V,\mu)}\label{eq:pn},
\end{equation}
where  the canonical partition function $Z(T,V,N)$ is obtained e.g. by
a projection of the grand partition function $\mathcal{Z}(T,V,\mu)$,
\begin{equation}\label{eq:projection}
 Z(T,V,N) = \frac{1}{2\pi}\int_{0}^{2\pi}d\left( \frac{\mu_I}{T}
					  \right)e^{-iN \frac{\mu_I}{T}
  }\mathcal{Z}(T,V,\mu = i\mu_I).
\end{equation}

In the HRG the probability distribution of the net-baryon number is,
within  the Boltzmann approximation,   given by the  Skellam function
\cite{skellam46:_frequen_distr_of_differ_between,braun-munzinger11:_net_proton_probab_distr_in}
  \begin{equation}
 P^{\text{S}}(N) = \left(\frac{b}{\bar{b}}\right)^{N/2}I_N(2\sqrt{b\bar{b}})e^{-(b+\bar{b})}\label{eq:pn_skellam}
\end{equation}
where $b = \langle N_b \rangle$ and $\bar{b} = \langle N_{\bar{b}}
\rangle$ are the thermal averages of the number of baryons and anti-baryons,
respectively.

The HRG model reproduces the particle yields in heavy
ion collisions in a broad energy range from SIS to LHC. Furthermore, it
describes the equation of state obtained in LQCD,
as well as the first and second order cumulants of the net baryon number
for temperatures below the chiral cross\-over temperature. On the other hand, as suggested in
\cite{friman11:_fluct_as_probe_of_qcd}, the {\em deviation} of higher order cumulants and their ratios
from the HRG results provides a potential signature for criticality at the phase boundary.

These considerations indicate, that the probability distribution of the HRG, the Skellam
function, offers an appropriate baseline for $P(N)$. Indeed, for small $N$, where
the probability distribution is fixed by the non-critical lowest order cumulants
the Skellam function provides a good approximation to $P(N)$. On the other hand, for
large $N$ the two distributions differ, since the critical fluctuations modify the tail of the distribution,
which in turn determines the higher cumulants.
Thus, it is natural to consider the Skellam function  as a reference for identifying
criticality in the probability distribution of the net-baryon number
\cite{braun-munzinger11:_net_proton_probab_distr_in}.
Specifically, we show that the ratio of $P(N)$ to the Skellam function exposes the
effect of critical fluctuations.

We extract the characteristic features of the probability distribution near
the chiral crossover transition within the $O(4)$ universality class by applying
the Functional Renormalization Group (FRG) approach to the  quark-meson (QM) model
\cite{wetterich93:_exact_flow_equat,berges02:_non_pertur_renor_flow_in,delamotte07}. The
QM model exhibits the relevant chiral symmetry of QCD,  and  belongs to the same $O(4)$ universality class
\cite{schaefer05:_phase_diagr_of_quark_meson_model,stokic10:_frg_scaling}.

The Lagrangian density in the QM model,  reads
\begin{gather}
 \mathcal{L}=\bar{q}[i\gamma_\mu \partial^\mu - g(\sigma + i\gamma_5
  \vec{\tau}\cdot\vec{\pi}) ]q +\frac{1}{2}(\partial_\mu
  \sigma)^2  +\frac{1}{2}(\partial_\mu \vec{\pi})^2 \nonumber\\
 -\frac{1}{2}m^2 \phi^2 + \frac{\lambda}{4}\phi^4 - h\sigma, 
 \label{eq:QMmodel}
\end{gather}
where $q$ and $\bar{q}$ denote the quark and anti-quark field coupled
with the $O(4)$ chiral meson multiplet
$\phi=(\sigma,\vec{\pi})$. The last three  terms in  Eq.~\eqref{eq:QMmodel}
constitute  the mesonic potential with the symmetry breaking term.

The thermodynamic potential is calculated in the QM model, within the
FRG approach \cite{wetterich93:_exact_flow_equat}.
Applying the optimized regulator to the exact flow
equation for the effective average action in the local potential
approximation \cite{wetterich93:_exact_flow_equat},
the flow equation for the scale dependent thermodynamic potential
density $\Omega_k$,  reads \cite{skokov10:_meson_fluct_and_therm_of}
\begin{gather}
 \partial_k \Omega_k(\rho)=\frac{k^4}{12\pi^2}\left[
 \frac{3}{E_\pi}\left\{1+2n_B(E_\pi)\right\}
	+\frac{1}{E_\sigma}\left\{1+2n_B(E_\sigma)\right\} \right. \nonumber\\
 \left.-\frac{24}{E_q}\left\{
 1-n_F(E_q^+)-n_F(E_q^-)\right\}
 \right]\label{eq:floweq}, 
\end{gather}
where $\rho=(\sigma^2+\vec{\pi}^2)/2$ is the reduced field variable and
$n_F$ and $n_B$ are the Fermi and Bose distribution functions,
respectively. The single particle
energies of $\pi$, $\sigma$ and $q$ are given by:
$E_\pi=\sqrt{k^2 + \bar{\Omega}^\prime_k},E_\sigma = \sqrt{k^2 +
 \bar{\Omega}^\prime_k + 2\rho \bar{\Omega}^{\prime\prime}_k}$,  and
$ E_q^\pm = \sqrt{k^2 + 2g^2 \rho}\pm \mu$,
where $\bar{\Omega}_k^{\prime}$ and
$\bar{\Omega}_k^{\prime\prime}$ denote the first and the second derivatives of
$\bar{\Omega}_k=\Omega_k+h\sqrt{2\rho_k}$, with respect to $\rho$.

The full thermodynamic potential is given by the minimum of
$\Omega_{k\rightarrow 0}(\rho)$.
We solve the flow equation \eqref{eq:floweq} numerically by making use
of  the Taylor expansion method
\cite{stokic,skokov10:_meson_fluct_and_therm_of}. At the ultraviolet
cutoff scale $k=\Lambda=1$ GeV, the initial
condition $\Omega_\Lambda(\rho)$ is fixed so as to reproduce the physical pion mass
$m_\pi=135$ MeV, and the sigma mass $m_\sigma = 640$ MeV. The strength of
the Yukawa coupling is fixed to be $g=3.2$ by the constituent quark
mass $M_q(T=\mu=0) = g\sigma_{k=0}(T=\mu=0)  = 300$ MeV,  with
$\sigma_{k=0}(T=\mu=0)$ $=$ $f_\pi =93$ MeV.

To avoid the  unphysical behavior of thermodynamic quantities at high
temperatures, we include the higher momentum contributions, beyond   the
cutoff scale $\Lambda$, by accounting for the $\mu-$ and $T-$dependent
thermodynamic potential obtained through the flow equation for a
non-interacting gas of  quarks and gluons
\cite{Braun_PRD70,skokov10:_meson_fluct_and_therm_of}.
The probability distribution $P(N)$  is then obtained from Eqs.~\eqref{eq:pn} and \eqref{eq:projection}, 
%by putting the obtained thermodynamic potential into
with $\mathcal{Z}=\exp({-\Omega_{k\rightarrow 0}/T})$.

In Ref.\ \cite{morita13:_net}, the FRG approach was applied to compute 
$P(N)$ of  the net-baryon number  within  the QM model at $\mu=0$.
In the present paper, we extend these studies to finite chemical potential and identify the qualitative structures
of the net baryon probability distribution which are due to $O(4)$ criticality at the chiral crossover transition.
We also evaluate the ratio of the data on the net proton probability distribution \cite{STAR_pn_2013}
to the Skellam function and discuss the results in the light of the theoretical considerations.

\begin{figure*}[!t]
 \includegraphics[width=3.375in]{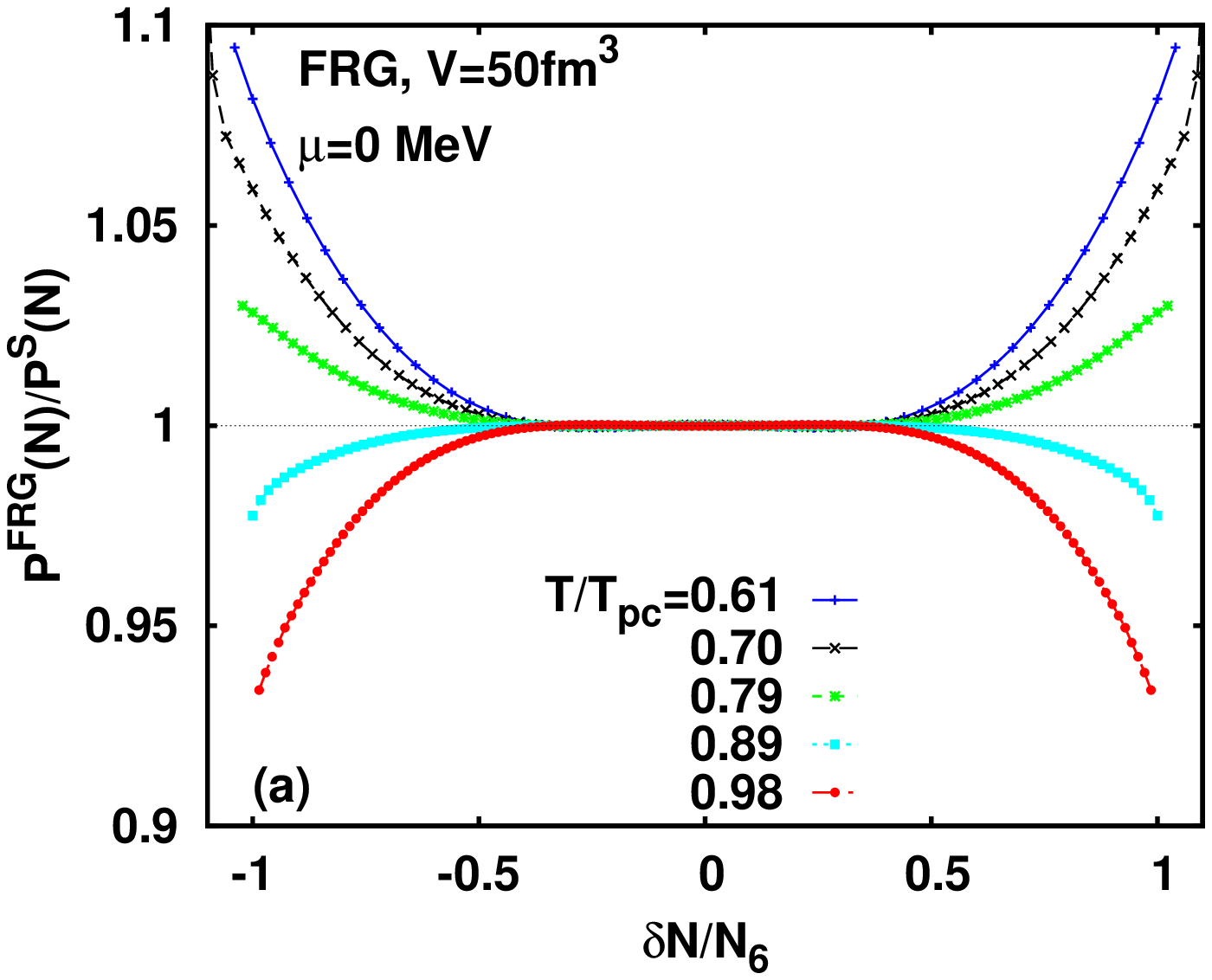}
 \includegraphics[width=3.375in]{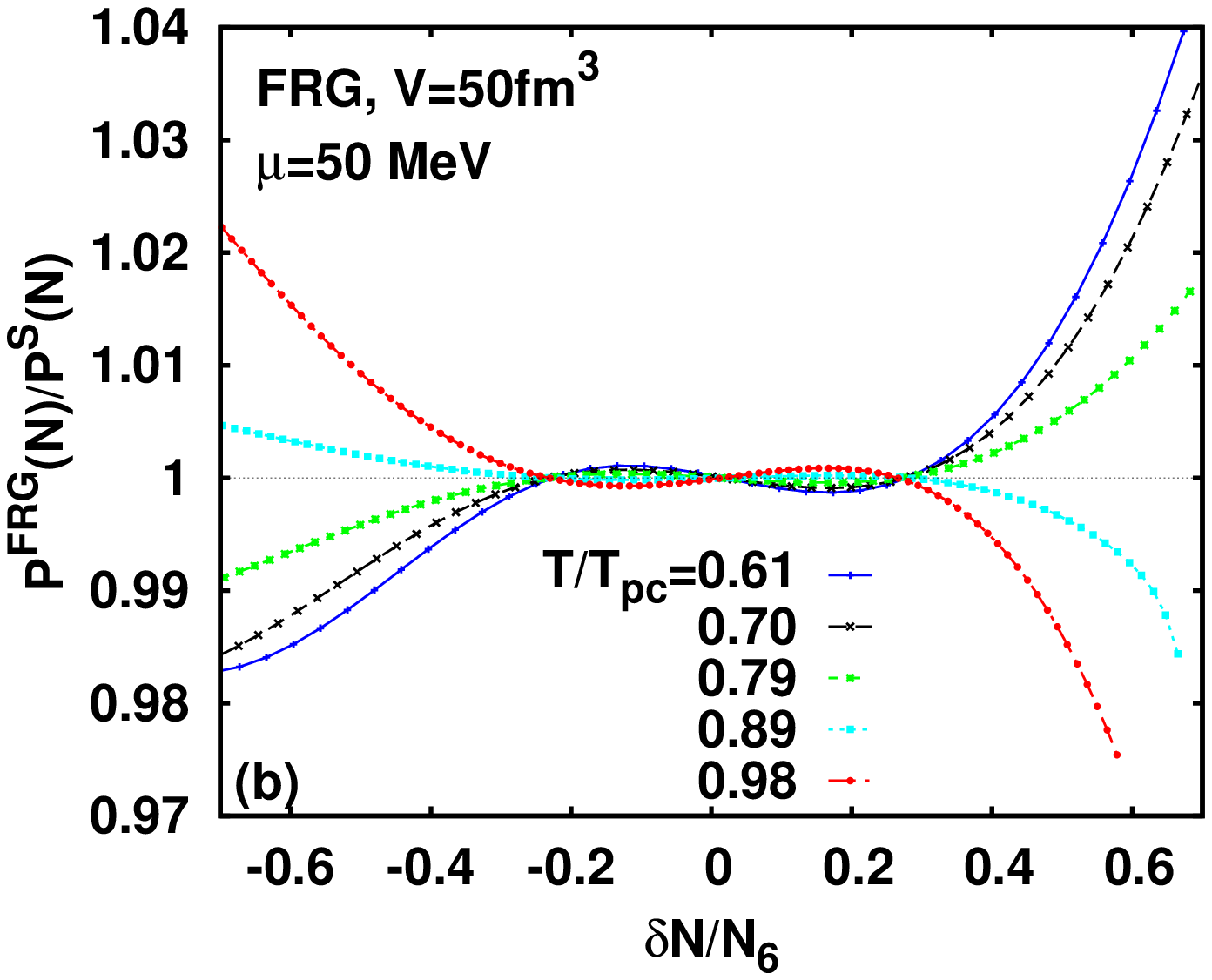}
  \caption{The ratio of the probability distribution obtained in
the  quark-meson model  $P^{\text{FRG}}(N)$ and the  Skellam
 distribution $P^S(N)$ with the same mean and variance as
 $P^{\text{FRG}}(N)$. In the left panel (a) shows the ratio  at
 $\mu=0$ for different  temperatures  $T/T_{pc}$, expressed  in units of
 the pseudocritical temperature $T_{pc}$, while in the right panel (b) shows the
 same ratio at  $\mu=50$ MeV. The quantities $\delta N$  and  $N_6$ are
 introduced in the text. }
 \label{fig:frg}
\end{figure*}

In general, the probability distribution $P(N)$ depends on the volume
parameter. However, as shown
in~\cite{morita13:_net},
the  volume dependence  of the re-scaled distribution $\sqrt V
P(N/\sqrt{V})$   is
st\-ro\-ngly reduced. This approximate  scaling is valid for both the
Skellam function, and
the $P(N)$ calculated within  the QM model for sufficiently large
$VT^3$, and is exact for a Gaussian distribution. Thus, in the
ratios of $P(N)$ and Skellam,  the leading volume dependence is
cancelled.
At finite density where $M >0$, the scaling property holds
for $\delta N = N-M$, after shifting the mean.  

In order to compute the cumulants $\chi_n$ reliably, knowledge of the
probability distribution $P(N)$ for sufficiently large $|\delta N|=|N-M|$ is needed. For a given order $n$,
it is sufficient to know the distribution $P(N)$ for $|\delta N|\leq N_{n}$,
where $N_{n}$ grows with $n$ and with the volume of the system \cite{morita13:_net}.
In the $O(4)$ universality class  and at $\mu=0$,  $\chi_6$ is the
first  cumulant which exhibits  criticality. Thus, to be able to identify
criticality in the  distribution,  we need to know $P(N)$ in the range needed to obtain a converged
result for $\chi_{6}$, i.e. for $|\delta N|< N_{6}$.  For a given volume,  $N_{6}$ is determined by
requiring that the sixth cumulant of the Skellam function is reproduced. As expected, we find that $N_{6}$ to a good approximation
scales with $\sqrt{V}$. Thus, in a plot of the ratio of $P(N)$ to the Skellam function as a
function of $\delta N/N_{6}$,  criticality is characterized by deviations from unity for $|\delta N/N_{6}|\lesssim 1$.

In Figure \ref{fig:frg}(a) we show  the ratio of $P^{FRG}(N)$,  computed
in the quark-meson model  at $\mu=0$  within the FRG approach
\cite{morita13:_net},  and  the Skellam distribution  $P^S(N)$,
with the same variance, as a function of  $\delta N/N_6$. The
results are shown for different temperatures $T/T_{pc}$,  where $T_{pc}$
is the chiral crossover or pseudocritical temperature.

This ratio exhibits a characteristic dependence on temperature, as
$T_{pc}$ is approached from below.
For  $T\simeq T_{pc}$, the ratio $P^{FRG}(N)/P^S(N)$ is less than unity,
indicating a narrowing of the probability distribution for larger $|\delta N|$, owing to $O(4)$  criticality.
Indeed, the decrease  of the probability ratio for $\delta N/N_{6}\simeq 1$
near $T_{pc}$ is responsible for the negative values of $\chi_6$, which
are characteristic of the chiral crossover transition in the $O(4)$ universality
class \cite{morita13:_net}. We note,  that the narrowing of $P^{(FRG)}(N)$ for smaller values of $\delta
N/N_6$,  seen in Fig. \ref{fig:frg}(a), can be
partly  attributed to the non-critical reduction of $\chi_4$.  However, the
smoothly decreasing tail of $P(N)$,  close to
$|\delta N / N_6| \lesssim 1$,  is entirely  due to
$O(4)$  criticality, resulting in the characteristic shape and the
negative structure of $\chi_6$.
Consequently, shrinking probability distribution, relative to the
Skellam one, can indeed be considered  as a necessary condition   for
$O(4)$ criticality \cite{morita13:_net}.

At finite chemical potential, the probability distribution $P(N)$  of
the net-baryon number,  is no longer  symmetric around the mean. Thus,
it is not a priori clear,  how the distribution is modified by  $O(4)$ criticality.

The asymmetry of $P(N)$  at $\mu\neq 0$,  appears due to the fugacity
factor $e^{\mu N/T}$ in Eq.~\eqref{eq:pn},   which  suppresses the
contribution from $ N < 0$ and enhances that from $ N > 0$.
Consequently,  at finite chemical potential,  the  tail of the probability distribution $P(N)$
is enhanced, and criticality is expected to appear at smaller
$|\delta N|$, and thus also in lower order cumulants.

Figure \ref{fig:frg}(b) shows the ratio $P^{FRG}(N)/P^S(N)$  obtained in the
QM model at  $\mu=50$ MeV.
Below the pseudo critical temperature $T_{pc}$, the distribution is asymmetric,
with an enhanced  tail relative to the Skellam function for positive and a
suppressed tail for negative values of $\delta N$. However, as $T_{pc}$ is approached, there is a
qualitative change of the properties of the distribution, resulting in a
narrowing for positive $\delta N$. Moreover, a comparison of
Figs.~\ref{fig:frg}(a) and (b) shows that at finite $\mu$ the narrowing
of $P(N)$ begins at smaller values of $\delta N/N_6 $.
The stronger narrowing of the distribution is consistent with the fact that
at finite $\mu$, already the third cumulant exhibits $O(4)$ critical
behavior~\cite{friman11:_fluct_as_probe_of_qcd}.
On the other hand, for negative $\delta N$ the ratio of the distributions exhibits the opposite
behavior, reflecting the asymmetry of the probability distribution at non-zero net baryon density.

At finite chemical potential and at large $|\delta N|$,  the
calculations of $P(N)$ are difficult due to the oscillating nature
of the integrand in the projection on the canonical partition function (\ref{eq:projection}). The
numerical integration yields reliable results only up to $\delta N/N_6 \leq 0.6$.
Consequently, the complete  $\chi_6$ cannot be reconstructed
due to insufficient information on the tail of the distribution. Nevertheless, the
narrowing of the probability ratio shown in Fig.~\ref{fig:frg}(b)
clearly exhibits the characteristic features
of $P(N)$, which are due to $O(4)$ criticality. Evidently the deviation
of $P^{FRG}(N)/P^S(N)$  from unity near $T_{pc}(\mu)$ will grow with
increasing $|\delta N/N_6|$ and $\mu$.

 \begin{figure}[ht]
  \includegraphics[width=3.375in]{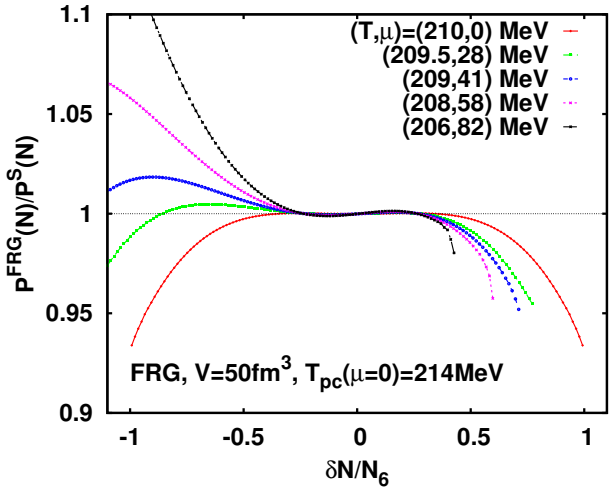}
  \caption{The probability ratio $P^{FRG}(N)/P^S(N)$ in a set of  points in the $(T,\mu)$ plane.
The points lie on an approximate freeze-out line, specified in the text. }
  \label{fig:pnratio_pb}
 \end{figure}

\section{O(4) criticality in heavy ion collisions}
In heavy ion collisions, particle yields, charge densities and their variance
are described consistently by the HRG model on the same chemical freeze-out line in the
$(T,\mu_B)$-plane
\cite{STAR_pn_2013,cleymans06:_compar,karsch11:_probin_freez_out_condit_in}.
For a given collision energy one can identify a unique point on the
freeze-out line. If the freeze-out takes place sufficiently  close to the
chiral crossover transition, the critical fluctuations are expected to leave a characteristic
imprint on the cumulants and on the corresponding probability distribution.

In Fig.~\ref{fig:pnratio_pb} we illustrate the expected structure of the probability distribution at
chemical freeze-out by showing the QM
model results for  $P^{\text{FRG}}(N)/P^S(N)$  at a set of
points in the $(T,\mu)$ plane. They lie on the approximate freeze-out line, defined by requiring the same
variance per unit volume of the net baryon number as in the $\mu=0$ point. The $\mu$ dependence of the
ratios, with a narrowing of the distribution for positive and a broadening for negative $\delta N$ with increasing $\mu$ is characteristic
for the critical region. As shown in Fig.\ \ref{fig:frg}, the distribution in a non-critical system
exhibits the opposite trend, with a broadening for positive and a narrowing for negative  $\delta N$.

In general, the measurement of higher order cumulants,\\
which are particularly sensitive
to criticality,  need high statistics owing to the increasing importance
of the tail of the distribution.
Furthermore,  the experimental conditions, such as acceptance corrections,
must be under control in order to make a meaningful comparison of the measured
cumulants and their probability distribution with
theoretical predictions \cite{luo11:_probin_qcd,luo12:_error_estim_for_momen_analy,chen13:_sixth_order_cumul_of_net}.

Recently the STAR Collaboration has published extensive results on the probability
distribution of the net-proton number $\Delta N_p = N_p-N_{\bar{p}}$ and
the corresponding cumulants up
to the fourth order,  obtained in heavy ion collisions~\cite{STAR_pn_2013}.
Also  preliminary results on the sixth
order cumulant have been  presented by STAR for several collision
energies and centralities~\cite{chen13:_sixth_order_cumul_of_net}.

While the cumulant ratios measured by STAR~\cite{STAR_pn_2013} were
efficiency  corrected and tested against possible modifications due to
volume fluctuations and accepted kinematical windows, the data on the probability
distributions of the net proton number are uncorrected.
Furthermore, in the model calculations, the net baryon number rather
than the net proton number measured by the STAR collaboration, is
considered. 
We assume, that the criticality due  to  chiral symmetry  restoration,
which appears in the net baryon number is also reflected in fluctuations
of the net proton number.  However, as shown in
Refs.~\cite{kitazawa04:_relat_between_baryon_number_fluct} and
\cite{Bluhm}, the quantitative differences between cumulants of the net
baryon and the net proton number are not excluded.  Therefore, the
significance of a direct comparison of model predictions with the
measured probability distribution is {\em a priori} not clear-cut.
Here we assume the isospin invariance such that the net baryon number
fluctuations are equivalent to those of net proton number and invoke the
generic properties of the probability distribution of the net baryon
number due to $O(4)$ criticality found in the QM model persist in the
net proton number probability distribution. 

Nevertheless, we have verified  that the data on $P(\Delta N_p)$ obtained by STAR
\cite{STAR_pn_2013} are dominated by physics. The contribution of volume
fluctuations to the data is small, as demonstrated by
the approximate scaling  of $P(\Delta N_p)$ with the standard
deviation $\sigma$ in central and  semi central collisions, for
$\sqrt{s_{NN}}\geq 19$GeV. This scaling
holds also for the Skellam and $P^{(FRG)}(N)$ distributions for sufficiently large volumes.
Moreover, the ratios of cumulants  computed directly from the
uncorrected $P(\Delta N_p)$ data \cite{STAR_pn_2013}, exhibit similar
systematics as the efficiency corrected ratios.
These tests indicate that the data may yield at least a qualitative
indication whether the measured distribution  exhibits criticality or not.

\begin{figure}[!t]
 \includegraphics[width=3.375in]{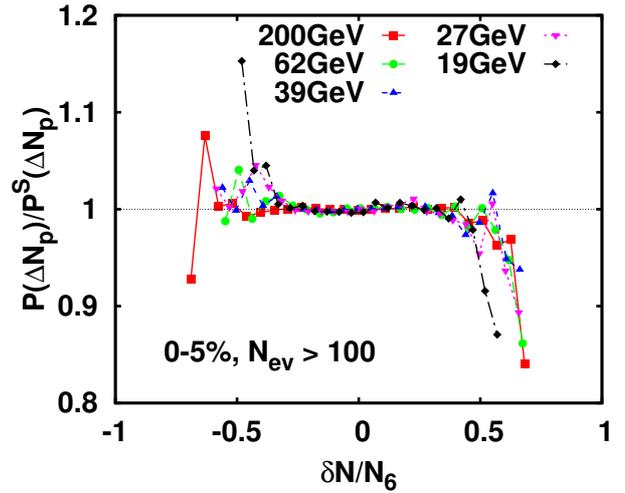}
 \caption{Ratios of the efficiency uncorrected probability distributions
 of the net-proton number $P(\Delta N_p)$ by STAR Collaboration
 \cite{STAR_pn_2013} to the Skellam function $P^S(\Delta N_p)$  with the same
 mean and variance as $P(\Delta N_p)$.
The data are for  the most central Au-Au collisions, with the number of
events $N_{\rm ev}>100$.  }
 \label{fig:ratio_STAR}

\end{figure}

In Fig.\ \ref{fig:ratio_STAR} we show the probability ratio
$P(\Delta N_p)/P^S(\Delta N_p)$
obtained from the uncorrected  data \cite{STAR_pn_2013} in the highest
centrality bin, for $\sqrt{s_{NN}}=200,62.4,39,27$ and 19.6GeV. The
probability ratio is constructed  using the same method as in
Fig.~\ref{fig:frg}. In order to avoid  large uncertainties, we have restricted
the data to those with more than 100 events. Consequently,
the probability distributions are limited to $|\delta N/N_6|<
0.5$.   This implies, that the present statistics does not allow for a
reliable estimate of the sixth order cumulant.
Nonetheless,  the ratios in  Fig.~\ref{fig:ratio_STAR} clearly exhibit a
structure qualitatively  similar to that shown in Figs.~\ref{fig:frg}  and
\ref{fig:pnratio_pb} which is a reflection of the underlying $O(4)$ criticality.
In particular, the characteristic narrowing  of the
probability distribution relative to the Skellam function for positive $\delta N$,  and  early drop of the ratio below
unity for $\sqrt{s} =19.6$GeV,  are characteristic signatures for $O(4)$  criticality at
non-zero chemical potential.

There are several potential contributions to the cumulants and the probability distribution from sources other than critical fluctuations
\cite{bzdak13:_baryon_number_conser_and_cumul,bzdak12:_accep_correc_to_net_baryon,kitazawa04:_relat_between_baryon_number_fluct,ono13:_effec_of_secon_proton_baryon},
as well as experimental issues e.g. regarding efficiency corrections
\cite{tang13:_effec_of_detec_effic_multip_distr}. Thus,
a final conclusion on the criticality of $P(\Delta N_p)$ can be drawn only
once the role of these effects has been sorted out.

In \cite{STAR_pn_2013}, the cumulant ratios $S\sigma$ and
$\kappa \sigma^2$ are analyzed with efficiency and centrality bin width
corrections. By constructing $S\sigma$ and $\kappa \sigma^2$ from the
uncorrected $P(N)$ data discussed above, we have found that the deviations
from Skellam distribution are slightly smaller than that seen  in the
corrected ratios. However the systematics  and the energy  dependence is
almost the same. Therefore, we regard the results shown in
Fig.~\ref{fig:ratio_STAR} as the lower limit for possible deviations
from the Skellam function.
We stress,  that the present method provides a transparent framework,
where such corrections can be included. If the narrowing of
$P(\Delta N_p)$ relative to  the Skellam, as seen in Fig. \ref{fig:ratio_STAR},  is
still observed after  these corrections are included, then this will
provide potential  evidence   for remnants of the chiral crossover transition in
experimental data.

\section{O(4) criticality and binomial distributions}
In order to reveal the $O(4)$ criticality in the net baryon number
probability distribution
and in the corresponding cumulants we have used the
Skellam distribution as a reference. The Skellam function is the
natural choice, since in heavy ion collisions  data on
particle yields, as well as the QCD thermodynamics,  are well
reproduced by the hadron resonance gas partition function. In the HRG,
baryons multiplicity is distributed according to Poisson and the net-charge
distribution is then given by the Skellam function.

However, it has been shown that the lowest cumulants of the net proton fluctuations and the corresponding probability
distributions obtained by the STAR Collaboration are
consistent also with negative binomial (NBD) or binomial (BD)
distributions  \cite{STAR_pn_2013,Westfall}. Thus, it is of interest to
verify to what extent these distributions can describe
critical fluctuations at the chiral transition. This study can be done within the QM model,  where
the  $O(4)$ critical structure of the cumulants and  the corresponding
probability distribution are manifest.

In the case of NBD or BD, the net-baryon probability  distribution  is
constructed assuming independent emission of baryons and antibaryons,
\begin{align}
 P^{NBD}(n;r,p)&= \frac{\Gamma(n+r)}{n!\Gamma(r)}p^n (1-p)^r,\\
 P^{BD}(n;r,p)&= \frac{\Gamma(r+1)}{n!\Gamma(r-n+1)}p^n (1-p)^{r-n},
\end{align}
with $n$ being the number of baryons or antibaryons.
For $\mu =0$, the property of the net-baryon $P(N)$ is uniquely
determined by two parameters $(r,p)$,  characterizing the
NBD or BD. Using the additive property of the cumulants, one finds
\begin{align}
 \chi_2^{\text{NBD}}&= \frac{2rp}{(1-p)^2},\label{NBD2}\\
 \chi_4^{\text{NBD}}&= \frac{2rp(6p+(1-p)^2)}{(1-p)^4},\\
 \chi_6^{\text{NBD}}&= \frac{2rp(1+26p+66p^2+26p^3+p^4)}{(1-p)^6},
\end{align}
and
\begin{align}
 \chi_2^{\text{BD}}&= 2rp(1-p),\\
 \chi_4^{\text{BD}}&= 2rp(1-p)(1-6p+6p^2),\\
 \chi_6^{\text{BD}}&= 2rp(1-p)(1-30p+150p^2-240p^3+120p^4).\label{BD6}
\end{align}
We construct the reference NBD/BD, which has the same $\chi_2$ and
$\chi_4$ as the $O(4)$ distribution obtained in the QM model within the
FRG method\footnote{We assume that the NBD/BD distribution provides an optimal description of the
non-critical observables, the baryon and antibaryon distributions as well as the first two non-zero cumulants.
In general, the situation is less favorable~\cite{Westfall}.}.

From Eqs.~\eqref{NBD2}-\eqref{BD6}, it is clear, that
$\chi_4/\chi_2 > 1$ for NBD and
$\chi_4 / \chi_2 < 1$ for BD. For $\chi_4/\chi_2=1$, both distributions
are reduced to the Skellam function for $r \rightarrow \infty$ and
$ p\rightarrow 0$.
Therefore, we use NBD for temperatures where $\chi_4/\chi_2 > 1$ and BD
for $\chi_4/\chi_2 < 1$. This implies, in particular, that NBD cannot describe
fluctuations around $T_{pc}$,  where $\chi_4/\chi_2 < 1$.

\begin{figure}[!t]
 \includegraphics[width=3.375in]{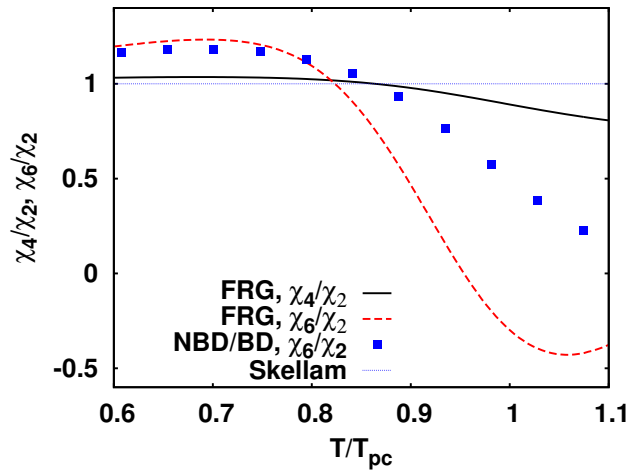}
 \caption{Ratios of cumulants computed in the quark-meson model at $\mu=0$
 within the FRG approach. Also shown is the  $\chi_6/\chi_2$ ratio
 obtained from the binomial (BD) and negative binomial (NBD)
 distributions with parameters fixed so as  to reproduce the model results
 for $\chi_2$ and $\chi_4$. The temperature is normalized to the
 pseudocritical temperature $T_{pc}$.   }
 \label{fig:nb_cum}
\end{figure}

Figure~\ref{fig:nb_cum} shows the ratios $\chi_{4}/\chi_{2}$ and $\chi_{6}/\chi_{2}$ obtained in
the QM model as functions of temperature near $T_{pc}$ and at
$\mu=0$. Also shown are the corresponding ratios obtained
from the NBD/BD distributions, where the parameters  $(r,p)$  are fixed so as to reproduce
$\chi_2$ and $\chi_4$.   Clearly the ratio
$\chi_6/\chi_2$, in particular the negative values near $T_{pc}$,
are not reproduced by the binomial distribution. We therefore conclude that the NBD/BD
distributions clearly fail to describe the critical fluctuations near the chiral transition.

At $\mu\neq 0$, the baryon and antibaryon distributions are different,
so that there are two sets of parameters $(r,p)$ in the NBD/BD, one
for protons and one for antiprotons. Thus, in principle, one may construct a reference distribution for
the net baryon number, $P(N)$, which reproduces the four leading cumulants\footnote{Note,
however, that $-1/2 \leq  \chi_4/\chi_2 < 1$ and
$-7/8 \leq \chi_6/\chi_2 < 1$ in BD.}, i.e. $\chi_n$ with  $n=1,2,3,4$, ignoring the fact
that at non-zero $\mu$, $\chi_{3}$ and $\chi_{4}$ may be affected by
criticality. We stress that it is unclear whether these parameters also yield
a good description of the baryon and antibaryon distributions.
Moreover, also in this case, the NBD/BD distributions that reproduce the leading cumulants, cannot describe
higher order ones, $\chi_n$ with  $n > 4$, nor the
tail of distribution, $P(N)$. Thus, an unambiguous verification of critical fluctuations in heavy ion collisions,
requires knowledge of the tail of net proton distribution $P(N)$, so that
the sixth order cumulant can be determined reliably, although at non-zero $\mu$, an indication of criticality may
be exposed in cumulants of lower order.

\section{Concluding remarks}
We have discussed the properties  of the net-baryon number probability
distribution $P(N)$ near the chiral crossover transition at vanishing and at finite
baryon  chemical potential. The critical properties  of
$P(N)$ in the quark-meson model were obtained within the functional renormalization
group approach. In the chiral limit this model  exhibits a second
order phase transition belonging to $O(4)$ universality
class in three dimensions.

At the physical value of the pion mass, the $O(4)$  criticality
is reflected in the tail of the  distribution $P(N)$. We have shown that
the ratio of $P(N)$ to the Skellam function $P^S(N)$,  constructed with
the same mean $M$ and variance as $P(N)$, clearly exhibits the influence
of the $O(4)$ criticality on the probability distribution.

We have shown that at vanishing chemical potential there is a
characteristic reduction  of this ratio below unity near
the phase boundary. At finite chemical potential,
the ratio
$P(N)$ $/$ $P^S(N)$ exhibits a characteristic
asymmetry in $\delta N=N-M$.  For $\delta N < 0$,   the
probability ratio is enhanced near the $O(4)$ pseudocritical point, while for $\delta N > 0$
it is suppressed. The asymmetry of the distribution is enhanced with increasing
$\mu$ along the freeze-out line.

The relevance  of our  results  for heavy ion experiments was
discussed. In particular,  we have computed the corresponding
probability ratios for
the efficiency uncorrected
net proton number obtained by the STAR Collaboration, and discussed their interpretation.
%We found that the presently available data show a similar trend to the
%$O(4)$ criticality. However, since they were not corrected for
%efficiency and other non-critical sources of fluctuation may contribute,
%the conclusions must be drawn after these issues are clarified.
%do not provide a unique signature for critical fluctuations
%at the chiral transition.
We have also demonstrated that $O(4)$ criticality, in particular its reflection in higher cumulants
of the net baryon number, is not consistent with a description of the baryon and antibaryon multiplicities in terms of
binomial  or negative-binomial distributions.

Finally we stress that an unambiguous identification of $O(4)$ chiral
criticality on the phase boundary, requires high statistics data on
the net proton probability distribution over a range in $\delta N$ that
allows a reliable determination of the sixth order cumulant.

\section*{Acknowledgement}
We thank X.~Luo and Nu ~Xu for valuable discussions and for  providing
numerical data  of the STAR Collaboration. We also acknowledge
stimulating discussions with  P. Braun-Mun\-zin\-ger, F. Karsch,
V. Koch, A. Ohnishi, C. Sasaki and V. Skokov.
We acknowledge the support of EMMI for the EMMI Rapid Reaction Task
Force on ``Probing the Phase Structure of Strongly Interacting
Matter with Fluctuations'' where part of this work was done.
KM and KR acknowledge the Yukawa Institute for Theoretical Physics,
Kyoto University, where this work was developed during
the YITP Workshop YITP-T-13-05 on ``New Frontiers in QCD 2013''.
KM was supported by HIC for FAIR, the Yukawa  International Program for
Quark-Hadron Sciences at  Kyoto University, by the Grant-in-Aid for
Scientific Research from JSPS No.24540271 and by the Grant-in-Aid for
Scientific Research on Innovative Areas from MEXT (No.~24105008).
KR and KM acknowledge partial support of the Polish Science Foundation
(NCN), under Maestro grant 2013/10/A/ST2/00106. The work of BF is
supported in part by EMMI.


\section*{References}
\begin{thebibliography}{10}
\expandafter\ifx\csname url\endcsname\relax
  \def\url#1{\texttt{#1}}\fi
\expandafter\ifx\csname urlprefix\endcsname\relax\def\urlprefix{URL }\fi
\expandafter\ifx\csname href\endcsname\relax
  \def\href#1#2{#2} \def\path#1{#1}\fi

\bibitem{stephanov99:_event_by_event_fluct_in}
M.~Stephanov, K.~Rajagopal, E.~Shuryak, Phys. Rev. D 60 (1999) 114028.

\bibitem{asakawa00:_fluct_probes_of_quark_decon}
M.~Asakawa, U.~W. Heinz, B.~M\"{u}ller, Phys. Rev. Lett. 85 (2000) 2072.

\bibitem{jeon00:_charg_partic_ratio_fluct_as}
S.~Jeon, V.~Koch, Phys. Rev. Lett. 85 (2000) 2076.

\bibitem{hatta03:_proton_number_fluct_as_signal}
Y.~Hatta, M.~A. Stephanov, Phys. Rev. Lett. 91 (2003) 102003.

\bibitem{skokov10:_meson_fluct_and_therm_of}
V.~Skokov, B.~Stokic, B.~Friman, K.~Redlich, Phys. Rev. C 82 (2010) 015206.

\bibitem{stephanov09:_non_gauss_fluct_near_qcd_critic_point}
M.~A. Stephanov, Phys. Rev. Lett. 102 (2009) 032301.

\bibitem{stephanov11:_sign_of_kurtos_near_qcd_critic_point}
M.~A. Stephanov, Phys. Rev. Lett. 107 (2011) 052301.

\bibitem{asakawa89:_chiral_restor_at_finit_densit_and_temper}
M.~Asakawa, K.~Yazaki, Nucl. Phys. A504 (1989) 668.

\bibitem{sasaki08:_chiral_phase_trans_in_presen}
C.~Sasaki, F.~Friman, K.~Redlich, Phys. Rev. D 77 (2008) 034024.

\bibitem{forcrand09:_simul_qcd_at_finit_densit}
P.~de~Forcrand, Proc. Sci. LAT2009 (2009) 010.

\bibitem{forcrand07:_n_qcd}
P.~de~Forcrand, O.~Philipsen, JHEP 0701 (2007) 077.

\bibitem{pisarski84:_remar_chiral_trans_in_chrom}
R.~D. Pisarski, F.~Wilczek, Phys. Rev. D 29 (1984) 338.

\bibitem{ejiri09:_magnet_equat_of_state_in_flavor_qcd}
S.~Ejiri, F.~Karsch, E.~Laermann, C.~Miao, S.~Mukherjee, P.~Petreczky,
  C.~Schmidt, W.~Soeldner, W.~Unger, Phys. Rev. D 80 (2009) 094505.

\bibitem{karsch11:_probin_freez_out_condit_in}
F.~Karsch, K.~Redlich, Phys. Lett. B 695 (2011) 136.

\bibitem{Statmodelreview_QGP3}
P.~Braun-Munzinger, K.~Redlich, J.~Stachel, in: R.~C. Hwa, X.~N. Wang (Eds.),
  Quark-Gluon Plasma 3, World Scientific, 2004.

\bibitem{borsanyi10:_qcd}
S.~Bors\'{a}nyi, G.~Endr\"{o}di, Z.~Fodor, A.~Jakov\'{a}c, S.~D. Katz,
  S.~Krieg, C.~Ratti, K.~K. Szab\'{o}, JHEP 1011 (2010) 077.
%\newblock \href {http://arxiv.org/abs/1007.2580} {\path{arXiv:1007.2580}}.

\bibitem{huovinen10:_qcd}
P.~Huovinen, P.~Petreczky, Nucl. Phys. A837 (2010) 26.

\bibitem{friman11:_fluct_as_probe_of_qcd}
B.~Friman, F.~Karsch, K.~Redlich, V.~Skokov, Eur. Phys. J. C 71 (2011) 1694.

\bibitem{aggarwal10:_higher_momen_of_net_proton}
M.~M. Aggarwal, et~al., Phys. Rev. Lett. 105 (2010) 022302.

\bibitem{luo13:_searc_for_qcd_critic_point}
X.~Luo, Nucl. Phys. A904-905 (2013) 911c.

\bibitem{braun-munzinger11:_net_proton_probab_distr_in}
P.~Braun-Munzinger, B.~Friman, F.~Karsch, K.~Redlich, V.~Skokov, Phys. Rev. C
  84 (2011) 064911.

\bibitem{STAR_pn_2013}
L.~Adamczyk, et~al., Phys. Rev. Lett. 112  (2014) 032302.
%\newblock \href {http://arxiv.org/abs/arXiv:1309.5681v1 [nucl-ex]}
%  {\path{arXiv:1309.5681v1 [nucl-ex]}}.

\bibitem{morita12:_baryon_number_probab_distr_near}
K.~Morita, V.~Skokov, B.~Friman, K.~Redlich, Eur. Phys. J. C 74 (2013) 2706.
%\newblock \href {http://arxiv.org/abs/1211.4703v2} {\path{arXiv:1211.4703v2}}.

\bibitem{morita13:_net}
K.~Morita, B.~Friman, K.~Redlich, V.~Skokov, Phys. Rev. C 88 (2013) 034903.
%  1301.2873.

\bibitem{skellam46:_frequen_distr_of_differ_between}
J.~G. Skellam, Journal of the Royal Statistical Society Series A 109 (1946) 3.

\bibitem{wetterich93:_exact_flow_equat}
C.~Wetterich, Phys. Lett. B 301 (1993) 90.

\bibitem{berges02:_non_pertur_renor_flow_in}
J.~Berges, N.~Tetradis, C.~Wetterich, Phys. Rept 363 (2002) 223.

\bibitem{delamotte07}
B.~Delamotte (2007).
\newblock \href {http://arxiv.org/abs/cond-mat/0702365}
  {\path{arXiv:cond-mat/0702365}}.

\bibitem{schaefer05:_phase_diagr_of_quark_meson_model}
B.~J. Schaefer, J.~Wambach, Nucl. Phys. 757 (2005) 479.

\bibitem{stokic10:_frg_scaling}
B.~Stoki\'{c}, B.~Friman, K.~Redlich, Eur. Phys. J. C 67 (2010) 425.
\bibitem{stokic}
 B.~Stokic, B.~Friman and K.~Redlich,
  %``Kurtosis and compressibility near the chiral phase transition,''
  Phys.\ Lett.\ B {\bf 673}, 192 (2009).

\bibitem{Braun_PRD70}
	J.~Braun, H.~-J.~Pirner, K.~Schwenzer, Phys. Rev. D 70 (2004)
	085016.

\bibitem{cleymans06:_compar}
J.~Cleymans, H.~Oeschler, K.~Redlich, S.~Wheaton, Phys. Rev. C 73 (2006)
  034905.

\bibitem{luo11:_probin_qcd}
X.~Luo, J. Phys.: Conf. Ser. 316 (2011) 012003.

\bibitem{luo12:_error_estim_for_momen_analy}
X.~Luo, J. Phys. G: Nucl. Part. Phys. 39 (2012) 025008.

\bibitem{chen13:_sixth_order_cumul_of_net}
L.~Chen, Nucl. Phys. A904-905 (2013) 471c.

\bibitem{kitazawa04:_relat_between_baryon_number_fluct}
M.~Kitazawa, M.~Asakawa, Phys. Rev. C 86 (2012) 024904.

\bibitem{Bluhm}
M. Nahrgang, M. Bluhm, P. Alba, R. Bellwied, C. Ratti, arXiv:1402.1238.

\bibitem{bzdak13:_baryon_number_conser_and_cumul}
A.~Bzdak, V.~Koch, V.~Skokov, Phys. Rev. C 87 (2013) 014901.

\bibitem{bzdak12:_accep_correc_to_net_baryon}
A.~Bzdak, V.~Koch, Phys. Rev. C 86 (2012) 044904.



\bibitem{ono13:_effec_of_secon_proton_baryon}
H.~Ono, M.~Asawaka, M.~Kitazawa, Phys. Rev. C 87 (2013) 041901(R).

\bibitem{tang13:_effec_of_detec_effic_multip_distr}
A.~H. Tang, G.~Wang, Phys. Rev. C 88 (2013) 024905.

\bibitem{Westfall}T.~J. Tarnowsky and G.~D. Westfall, Phys. Lett. B 724
	(2013) 51.
\end{thebibliography}
\end{document}